\documentclass{article}

\usepackage{amsmath, amsthm, amssymb}
\usepackage{algorithm}
\usepackage{algpseudocode}
\usepackage{graphicx}
\usepackage{verbatim}
\usepackage{natbib}
\usepackage{caption}
\usepackage{subcaption}
\usepackage{fancyvrb}
\usepackage{enumerate}
\usepackage{enumitem}
\usepackage{relsize}
\usepackage{hyperref}
\usepackage[margin=1.5in]{geometry}
\hypersetup{colorlinks,citecolor=blue,urlcolor=blue,linkcolor=blue}
\usepackage{diagbox}
\usepackage{stefan_tex}
\usepackage{float}
\usepackage{url}
\usepackage{multirow}

\usepackage[utf8]{inputenc}
\usepackage[english]{babel}

\usepackage{accents}
\theoremstyle{definition}
\newtheorem{definition}{Definition}[section]

\usepackage{pgfplots}
\usepackage{tikzscale}
\pgfplotsset{compat=newest}
\usepgfplotslibrary{groupplots}
\usepgfplotslibrary{polar}
\usepgfplotslibrary{smithchart}
\usepgfplotslibrary{statistics}
\usepgfplotslibrary{dateplot}
\usepgfplotslibrary{external}
\usetikzlibrary{arrows.meta}
\usetikzlibrary{backgrounds}
\usepgfplotslibrary{patchplots}
\usepgfplotslibrary{fillbetween}
\pgfplotsset{%
layers/standard/.define layer set={%
    background,axis background,axis grid,axis ticks,axis lines,axis tick labels,pre main,main,axis descriptions,axis foreground%
}{grid style= {/pgfplots/on layer=axis grid},%
    tick style= {/pgfplots/on layer=axis ticks},%
    axis line style= {/pgfplots/on layer=axis lines},%
    label style= {/pgfplots/on layer=axis descriptions},%
    legend style= {/pgfplots/on layer=axis descriptions},%
    title style= {/pgfplots/on layer=axis descriptions},%
    colorbar style= {/pgfplots/on layer=axis descriptions},%
    ticklabel style= {/pgfplots/on layer=axis tick labels},%
    axis background@ style={/pgfplots/on layer=axis background},%
    3d box foreground style={/pgfplots/on layer=axis foreground},%
    },
}

\usepackage{soul}

\graphicspath{{./figures/}}

\theoremstyle{plain}

\theoremstyle{definition}

\newtheorem{assu}{Assumption}

\theoremstyle{remark}

\newtheorem{rema}{Remark}

\title{Partial Likelihood Thompson Sampling}

\author{
\makebox[45mm]{Han Wu} \\ Stanford University \and
\makebox[45mm]{Stefan Wager} \\  Stanford University}

\begin{document}

\maketitle

\begin{abstract}
  We consider the problem of deciding how best to target and prioritize existing vaccines that may offer protection against new variants of an infectious disease. Sequential experiments are a promising approach; however, challenges due to delayed feedback and the overall ebb and flow of disease prevalence make available methods inapplicable for this task. We present a method, partial likelihood Thompson sampling, that can handle these challenges. Our method involves running Thompson sampling with belief updates determined by partial likelihood each time we observe an event. To test our approach, we ran a semi-synthetic experiment based on 200 days of COVID-19 infection data in the US.
\end{abstract}

\section{Introduction}\label{sec:intro}
Methods for sequential experimentation have proven themselves as powerful and
versatile in a number of application areas, ranging from online advertising \citep{chapelle2011empirical}
and revenue management \citep{ferreira2018online} to website optimization
\citep{letham2019constrained}. These methods enable us to efficiently
optimize an explore-exploit tradeoff between first discovering which of a number of actions is best
and then efficiently deploying it once we've identified it.
One simple yet successful idea for doing so is Thompson sampling \citep{thompson1933likelihood,russo2018tutorial},
where an agent dynamically updates a belief distribution for the probability that each action they could take is best,
and then takes actions with propensity proportional to these beliefs.

An important and potentially promising application for sequential learning\footnote{In this paper, we use the terms sequential learning and sequential experimentation interchangeably.}
is in deciding how best to use existing vaccines to target new variants of an infectious
disease. For example, in the case of the COVID-19 pandemic, a number of vaccines were developed
and found to be safe and effective in protecting against the original viral strain; however,
new coronavirus variants then emerged that exhibited at least partial ability to evade protection from
vaccines, thus making it more difficult to contain the pandemic \citep{kustin2021evidence}.
In situations like this, it is of considerable value to promptly assess which of the existing
vaccines (if any) offer protection against the new variant.
\citet{castillo2021market} call for embedding vaccine trials on new COVID-19 variants within national vaccine rollouts,
and sequential learning seems like a perfect candidate for optimizing the resulting explore-exploit tradeoff across
vaccines. In this setup, we would only be comparing vaccines that have already been
established as generally safe and effective, and so the goal is to discover---as quickly as
possible---which vaccine is most effective in the context of interest.

The main difficulty in using sequential experiments for vaccine trials is that such trials involve delayed
feedback of a type that cannot readily be handled by available methods, including
Thompson sampling \citep{thompson1933likelihood,russo2018tutorial}
or the UCB algorithm \citep{lai1985asymptotically,Auer2004FinitetimeAO}.
The standard framework for sequential learning involves a tight feedback loop, where in
each time-step an agent chooses an action, sees the corresponding reward, and can then update their beliefs.
In a vaccine trial, however, we cannot immediately assess success after an innoculation;
rather, we can only wait and see whether the patient gets infected any time before the end of the trial.

There has been some recent work on adaptive trials with delayed feedback
\citep{grover2018best,joulani2013online,zhou2019learning};
however, available methods cannot simultaneously address some key difficulties that arise when testing
vaccines in a pandemic.
First, there is no (useful) upper bound on the delay separating an action and the corresponding reward.
Study subjects could experience a negative reward (i.e., get infected) anytime between when they're enrolled
in the study (and given a vaccine) and the end of the study.
Second, the rate of infections doesn't just depend on which vaccines are used, but also on the ebb and flow of the pandemic.
Any method that adaptively adjusts vaccine allocation frequencies without accounting for varying baseline
infection rates risks providing a biased comparison.
The main goal of this paper is to develop methods for sequential experimentation
that can handle the above challenges.

Our core proposal is to extend the Thompson sampling method for sequential Bayesian learning to the
proportional hazards model \citep{cox1972regression}, which is widely used in
medical statistics. In our context---as spelled out in more detail below---the proportional hazard model
posits that, at time $t$ and for an as-of-yet
uninfected person having received vaccine $k$, the instantaneous risk (i.e., hazard) of getting infected is of the form
$h_0(t) e^{-\theta_k}$. Here, $h_0(t)$ is the baseline hazard, i.e., the time-varying instantaneous risk that an unvaccinated
person gets infected, and $\theta_k$ captures the protective effect of the vaccine (the larger $\theta_k$ the better
the vaccine). The proportional hazards model is a natural fit for our setting in that it allows us to address
the challenges highlighted above (i.e., unbounded delays to observed infections and time-varying baseline
hazards), yet it has enough structure to enable sample-efficient learning.
One celebrated property of the proportional hazards model is that we can learn about the underlying
efficiency parameters $\theta_k$ via a partial likelihood in which
the baseline risk $h_0(\cdot)$ gets canceled out \citep{cox1975partial,efron1977efficiency}.

Our proposed approach, partial likelihood Thompson sampling (PLTS), involves running Thompson sampling with
belief updates determined by partial likelihood each time we observe an event (i.e., each time an already
vaccinated study participant gets infected). This differs from a Bernoulli bandit in that we do not control
when events may happen, and the relevant ``at risk'' sample size changes with time.
The resulting Bayesian problem doesn't have a closed-form solution for the
posterior, but we find the setting to be amenable to popular methods for approximate inference
with Thompson sampling---including Laplace approximation \citep{chapelle2011empirical,russo2018tutorial}.
While the use of partial likelihood
for Bayesian inference in general is well established \citep{kalbfleisch1978non}, we are not aware
of prior research on using proportional hazards modeling or partial likelihood for sequential experiments.

In a semi-synthetic study using data from the COVID-19 pandemic, we find that our approach can more reliably
identify the best vaccine than a classical randomized controlled trial (RCT), in which volunteers are assigned
to different treatments uniformly at random throughout the trial. Our approach also considerably reduces the
within-experiment regret from assigning study participants to sub-optimal vaccines.

\subsection{Related Work}

At a high level, sequential vaccine experiments can be seen as a bandit problem with partial,
delayed feedback: Feedback is partial because we only ever observe negative rewards (infections),
and delayed because it takes time for a study participant to potentially get infected post vaccination.
There is a large amount of work on bandits with full feedback:
\citet{dudik2011efficient} consider bandits with constant deterministic delays;
\citet{joulani2013online} study a setting where delays have bounded expectation;
\citet{Mandel_Liu_Brunskill_Popovic_2015} consider bounded delays in the stochastic multi-armed bandit problem;
\citet{Thune2019NonstochasticMB} work with bounded delays in the nonstochastic bandit problem.
Meanwhile, \citet{vernade2017stochastic} allow for partial feedback but assume i.i.d. delays with a known distribution,
and \citet{Manegueu2020StochasticBW} consider the same partially observable model but with the assumption that delay distributions
satisfy polynomial tail bounds.
\citet{MAB_unrestricted_delay} develop algorithms based on UCB and successive elimination that
allow for unrestricted delay distributions.  However, their bounds are vacuous in our scenario as we only observe infections
(i.e., negative rewards); and the delays considered in \citet{MAB_unrestricted_delay} are assumed to be i.i.d across time. 
Thus, we are not aware of existing methods studied in a setting that includes vaccine trials, where delays are unbounded and
time-varying, and positive rewards are never observed. We do note, however, that the method of \citet{Thune2019NonstochasticMB}
is one that---at least algorithmically---could be plausibly considered in our setting, and we use it as a baseline in our experiments;
see Section \ref{sec: experiments} for details.

Proportional hazards modeling and partial likelihood are core techniques in survival analysis.
\citet{cox1972regression} first proposed the proportional hazards model, while
\citet{cox1975partial} and \citet{efron1977efficiency} further developed
statistical theory for estimators based on partial likelihood. \citet{exact_method, breslow1974, efron1977efficiency} provided alternative
likelihood formulas when the event times are discrete with multiplicity. We also note a line of work justifying
the use of Bayesian methods on partial likelihood. \citet{kalbfleisch1978non} show that partial likelihood is a
limiting marginal posterior under noninformative priors for baseline hazards. \citet{10.1093/biomet/90.3.629}
further extend the result to scenarios with time-dependent covariates and time-varying regression parameters.
\citet{doi:https://doi.org/10.1002/9781118445112.stat06003} gives a comprehensive textbook treatment of Bayesian survival analysis.

Finally we note that using hazard rates to model the efficacy of vaccines is widely used in medical statistics;
see for example \citet{halloran1996}, \citet{halloran1998} and \citet{halloran1999}.
Thus, the main contribution of this paper is to leverage fundamental concepts in survival analysis and
classical vaccine RCTs, i.e., proportional hazards modeling and partial likelihood, to develop a new bandit algorithm
suitable for adaptive vaccine trials.

\section{Infection Modeling via Proportional Hazards}

We model vaccine trials as follows.
At the start of the trial (i.e., at time $t = 0$), some participants are recruited to
the trial and assigned to each vaccine group (arm) uniformly at random. After the initial assignment,
volunteers arrive over time, and we randomize and assign them to a treatment arm as soon as they
arrive. After enrollment, participants are followed until either they get infected or the
study ends; any infected participants are removed from the study at the moment they are infected.
Throughout, we use the following notation:
\begin{equation}
\begin{split}
&M_{t,k} = \text{\# participants assigned to arm $k$ by time $t$} \\
&m_{t,k} = \text{\# participants assigned to arm $k$ at time $t$} \\
&N_{t,k} = \text{\# observed infections in arm $k$ by time $t$} \\
&n_{t,k} = \text{\# observed infections in arm $k$ at time $t$} \\
& o_{t,k} = \text{\# participants remaining in arm $k$ at time $t$},
\end{split}
\end{equation}
i.e., $M_{t,k}$ and $N_{t,k}$ are cumulative sums of $m_{t,k}$ and $n_{t,k}$ respectively, and $o_{t,k} = M_{t,k} - N_{t,k}$.
We also denote the sum of these statistics across all arms as $M_t, m_t, N_t, n_t, o_t$. We also have the assumption that $n_{0,k} = 0$ for all $k$ since we do not observe any infection at the start of the trial and $m_{T,k} = 0$ for all $k$ since we do not assign any new participants when we end the trial.

This general model is formalized in Protocol \ref{protocol:trial}. One important case of this study
design is the batched setting we consider in this paper, where there are a finite number of time points participants can join the study and infections can be recorded. Specifically, at $t = 0,...., T$, we collect $m_t$ newly arrived participants and assign them to different groups and we also observe a vector of new infections $(n_{t,1},...,n_{t,K})$. At time $T$, we end the experiment. This is summarized in Protocol \ref{protocol:trial_discrete}. 

We model person-specific infection risk using the classical notion of a hazard rate,
as follows. We assume that each of the $k = 1, \, \ldots, \, K$ treatment arms is characterized
by a hazard rate $h_k(t)$, which captures the instantaneous risk that a person in study arm $k$
becomes infected at time $t$. Below, note that $o_{t,k}$ denotes the number of still
uninfected participants in arm $k$ at time $t$, and $h_k(t)$ describes the expected fraction
of these participants who will become infected in the next instant \citep{cox1984analysis}.

\begin{assu}
\label{assu:hazard}
For each study arm $k = 1, \, \ldots, \, K$, there is a hazard rate $h_k(t)$ such that,
for all $0 < t < T$,
\begin{equation}
h_k(t) = \lim_{dt \downarrow 0} \frac{1}{dt} \frac{\mathbb{E}_t[N_{t + dt, \, k} - N_{t, \, k}]}{o_{t,k}},
\end{equation}
where $\mathbb{E}_t$ denotes expectations conditionally on information available at time $t$.
\end{assu}

The key flexibility of Assumption \ref{assu:hazard} is that it allows infection risk to ebb and
flow over time: There may be some periods where very few people from any study arms are getting
infected, and others where infections are highly prevalent in some arms. However, Assumption 
\ref{assu:hazard} does impose non-trivial structure on the problem: For example, it implies that
the length of time a patient has been in the study does not affect their risk of getting infected.

Given Assumption \ref{assu:hazard}, \citet{halloran1999} defines vaccine efficiency in terms
of a ratio of hazard functions. Suppose that one of the study arms (without loss of generality
the first arm $k = 1$) is a placebo that does not provide any protection against infection. Then
the efficiency of the $k$-th vaccine depends on $h_k(t) / h_1(t)$.

\begin{definition}
Under Assumption \ref{assu:hazard},
for each non-placebo arm $k = 2, \, \ldots, \, K$,
the vaccine efficiency is
\begin{equation}
\text{VE}_k(t) = 1 - \frac{h_k(t)}{h_1(t)}.
\end{equation}
\end{definition}

 \setlength{\textfloatsep}{12pt}
 \begin{algorithm}[t]
    \floatname{algorithm}{Protocol}
    \caption{\label{protocol:trial} 
             General Vaccine Trial}
    \begin{algorithmic}
        \State \textbf{Input}: Length of experiment $T$, number of vaccines $K$
        \State Assign $m_0$ participants uniformly at $t = 0$
        \While{$t \leq T$}
            \If{$m_t \ne 0$}
            \State Assign $m_t$ participants to vaccine groups.
            \EndIf
            \State Observe a vector of infections $(n_{t,1},...,n_{t,K})$ and end trial for the infected participants.
        \EndWhile
    \end{algorithmic}
    % \vskip -0.1in
\end{algorithm}

 \setlength{\textfloatsep}{12pt}
 \begin{algorithm}[t]
    \floatname{algorithm}{Protocol}
    \caption{\label{protocol:trial_discrete} 
             Discrete Time Vaccine Trial}
    \begin{algorithmic}
        \State \textbf{Input}: Length of experiment $T$, number of vaccines $K$
        \For{$t = 0,1, ..., T$}
            \State Assign $m_t$ participants to vaccine groups.
            \State Observe a vector of infections $(n_{t,1},...,n_{t,K})$ and end trial for the infected participants.
        \EndFor
    \end{algorithmic}
    % \vskip -0.1in
\end{algorithm}

Given our assumptions so far, the vaccine efficiency $\text{VE}_k(t)$
may vary with time, which creates some potential ambiguity in defining
what the best vaccine is. Our next major assumption is that vaccine
efficiency doesn't change with time, i.e., equivalently, that the hazard
functions follow the proportional hazards model of \citet{cox1972regression}.

\begin{assu}
\label{assu:arm_prop_hazard}
For each study arm $k = 2, \, \ldots, \, K$, there is an efficiency parameter $\theta_k$ such that
\begin{equation}
\label{VE}
h_k(t) = h_1(t) e^{-\theta_k}, \ \ \ \ \text{VE}_k(t) = 1 - e^{-\theta_k}.
\end{equation}
\end{assu}

%\begin{assu}
%The placebo has a hazard rate $h_0(t)$, so $\theta_{\text{placebo}} = 0$.
%\end{assu}

Given Assumption \ref{assu:arm_prop_hazard}, the main task of interest in assessing vaccines' efficiency is to estimate the efficiency parameter $\theta_k$: The bigger the
$\theta_k$, the more effective the vaccine is. To illustrate this with a concrete
example, in initial studies, \citet{doi:10.1056/NEJMoa2034577} reported that the Pfizer COVID-19 vaccine
was 95\% effective in preventing infection while \citet{doi:10.1056/NEJMoa2035389} reported
that the Moderna COVID-19 vaccine was 94.1\% effective. In the context of our model, both
of these points estimates correspond to an efficiency parameter $\theta \approx 3$.  
\begin{rema}
We set $\theta_1 = 0$ so the placebo arm has the same hazard rate as the baseline hazard. 
\end{rema}
\begin{rema}
Here, for simplicity, we assume constant efficiency of the vaccine. We make this assumption because we are modeling the infection against a particular variant within the time span of the trial. So, for example, in studying COVID vaccine efficiency against the omicron variant, we would only count omicron infections as events, while ignoring infections with other variants (and $h_0(t)$ would be essentially 0 early in the pandemic until omicron got prevalent). Given the usual time of such vaccine trials (on the order of months) we think it is reasonable to consider a constant $\theta_k$. However, for longer experiments, it may be necessary to extend the model to allow for waning efficiency. We leave extensions to non-constant efficiency to future work.
%Here, for simplicity, we assume constant efficiency of the vaccine. This may be a reasonable assumption for experiments run on the order of months; however, for longer experiments, it may be necessary to extend the model to allow for waning effectiveness.
%We leave extensions to non-constant efficiency to future work.
\end{rema}

One major advantage of the proportional hazards model is that it enables a simple approach to
learning the efficiency parameters $\theta_k$ via partial likelihood
\citep{cox1972regression,cox1975partial,efron1977efficiency}, as follows.
Let us first suppose that the infection times (event times) of our participants in the trial are continuous as in Protocol \ref{protocol:trial} and recall at time $t$ the number of participants in each group is characterized by the vector $(o_{t,1}, ..., o_{t,K})$, i.e., this is the number of participants in each group who have joined the study, been assigned
a treatment, and have not yet been infected. Then, the conditional probability that a person in vaccine group $j$
is infected given that there is an infection at time $t$ is (let $\theta_1 = 0$):
\begin{align*}
& p_j(\theta_2,...,\theta_K\cond o_{t,1},....,o_{t,K}) \\
    & = \frac{h_0(t)e^{-\theta_j}}{\sum_{k=1}^{K}o_{t,k}h_0(t)e^{-\theta_k}}  = \frac{e^{-\theta_j}}{\sum_{k=1}^{K}o_{t,k}e^{-\theta_k}}.
\end{align*}
The unknown baseline hazard function cancels out because of the proportional hazards assumption.
Now suppose we have $J$ events (infections) happening at time $t_1 < t_2 < \cdots < t_J$ and event $j$ happened to group $I_j$.
We can then form the following partial likelihood,
\begin{equation}
\ell(\theta_1,...,\theta_K) = \prod_{j=1}^{J} \frac{e^{-\theta_{I_j}}}{\sum_{k=1}^{K}o_{t_j,k}e^{-\theta_k}}, \label{partial_lik}    
\end{equation}
which is a product over all the conditional probabilities of the observed events. It is a partial likelihood because we ignore all non-events. However, it is efficient for estimating the hazard rate parameters \citep{efron1977efficiency}. 

The partial likelihood \eqref{partial_lik} we obtained in the last section assumes continuous infections times where there are no ties in a single event time $t_j$. However, in Protocol \ref{protocol:trial_discrete} the event times will be $1,...,T$ and there could be multiple infections in a single vaccine group if the hazard rate is really high. Recall the definition of $(n_{t,1},...,n_{t,K})$ which denotes the number of infections happened in each vaccine group during the time interval $(t-1,t]$ and $n_t$ which denotes the sum of infections across all vaccine groups. In this case the exact likelihood proposed in \citet{cox1972regression} is the following 
\begin{equation}
  \ell(\theta_1,...,\theta_K)=  \prod_{t=1}^{T} \frac{e^{-\sum_{k=1}^{K}\theta_kn_{t,k}}}{\sum_{l \in R(n_t)} e^{-\theta(l)}}
\label{discrete_exact}
\end{equation}
where $R(n_t)$ is the set of all possible sets of $n_t$ participants from the risk set $(o_{t,1},..,o_{t,K})$ and $\theta(l)$ is the sum of all the $\theta$ values of the individuals in set $l$. Due to its complicated form, \citet{breslow1974} suggests using the following approximation
\begin{equation}
   \ell(\theta_1,...,\theta_K)= \prod_{t=1}^{T} \prod_{k=1}^{K} \left (\frac{e^{-\theta_k}}{\sum_{i=1}^{K}o_{t,i} e^{-\theta_i}} \right)^{n_{t,k}}
    \label{breslow_approx}
\end{equation}
of the exact partial likelihood \eqref{discrete_exact}, and this is what we do in our
approach.\footnote{Other approximations have also been proposed in case of ties,
notably that of \citet{efron1977efficiency}. Here, we use the Breslow approximation due to its simplicity
and the fact that in our experiments there are only a few ties---and so our results are not particularly
sensitive to the approximation method we use.}

\section{Partial Likelihood Thompson Sampling} \label{sec:plts}

In this section, we describe our proposed algorithm for sequential experimentation,
which we call Partial Likelihood Thompson Sampling (PLTS).
Thompson sampling \citep{thompson1933likelihood} is a Bayesian heuristic for sequential experiments that chooses the actions at each round according to the posterior probability that the action maximizes expected reward. This is usually implemented by sampling, where we sample an instance of environment from the posterior
and take the action that maximizes the expected reward \citep{russo2018tutorial}.

In our setting our model parameters are efficiency parameters $\theta_2,...\theta_K$ (recall we assume that $\theta_1 = 0$, i.e., that the first arm is a placebo). At each round we will get a sample of $(\theta_2, .., \theta_K)$ from the posterior and assign our participants accordingly. We start with uninformative prior for all the parameters as they are potentially unconstrained \citep{gelman2013bayesian}.
Then, following the blueprint of Thompson sampling (see Algorithm \ref{algo:PLTS}), we update the posterior
each time we collect new data; and here, we do so using the partial likelihood introduced in the previous section.
The use of partial likelihood for Bayesian posterior updates is further discussed in \citet{kalbfleisch1978non}.

Given this setup, it now remains to derive an efficient posterior sampling method for assigning new participants as they arrive. 
Since we put an uninformative prior on all the parameters $\theta_2,..., \theta_k$ the posterior at time $t$ given observed data $\mathcal{D}$ will be 
\begin{equation}
\begin{split}
     \mathbb{P}_t(\theta_2,..,\theta_K \cond \mathcal{D}) &=  \frac{p(\mathcal{D}\cond \theta_1,...,\theta_K)p(\theta_1,...,\theta_K)}{p(\mathcal{D})} \\
     & \propto \ell(\theta_1,...,\theta_K)   \\
   &= \prod_{l=1}^{t} \prod_{k=1}^{K} \left (\frac{e^{-\theta_k}}{\sum_{i=1}^{K}o_{l,i} e^{-\theta_i}} \right)^{n_{l,k}}.
\end{split}
   \label{eq:posterior}
\end{equation}

Now, one difficulty in using \eqref{eq:posterior} directly is that efficiently sampling from the posterior
is non-trivial. For computational tractability, we thus use the popular idea of replacing the exact posterior
with its Laplace approximation \citep{basu2020adaptive,chapelle2011empirical,gomezuribe2016online,russo2018tutorial}.
The main idea is as follows. Writing $ \mathbb{P}_t(\theta_2,..,\theta_K)$ for the partial likelihood used in \eqref{eq:posterior},
we see that ignoring constants,
\begin{align*}
  & \log\mathbb{P}_t(\theta_2,..,\theta_K\cond \mathcal{D}) \\
  & = \sum_{l=1}^{t} \left( -\sum_{k=1}^{K}n_{l,k} \theta_k - n_{l} \log\left(\sum_{i=1}^{K}o_{l,i} e^{-\theta_i}\right) \right) \\
  & = -\sum_{k=2}^{K}\sum_{l=1}^{t} n_{l,k}\theta_k - \sum_{l=1}^{t}n_{l}\log\left(o_{l,1} + \sum_{i=2}^{K}o_{l,i} e^{-\theta_i}\right).
\end{align*}
From the above formula we see that the logarithm of the likelihood is concave,
hence there exists a unique global maximizer of the likelihood. Laplace approximation
involves approximating the posterior with a Gaussian distribution centered at the posterior mode;
the inverse covariance matrix will be $-\nabla^2\log \mathbb{P}_t(\hat{\theta}_2,...,\hat{\theta}_K \cond \mathcal{D})$,
where $\hat{\theta}_2,...,\hat{\theta}_K$ are the posterior mode. Formally, this gives us an approximate sampling distribution
\begin{equation}
\label{maximizer}
\begin{split}
   & \hat{\theta}_2,....,\hat{\theta}_K = \argmax \mathbb{P}_t(\theta_2,...,\theta_K \cond\mathcal{D}) \\
   & \hat{\Sigma} = -\nabla^2\log \mathbb{P}_t(\hat{\theta}_2,...,\hat{\theta}_K\cond \mathcal{D}) \\
   & \mathbb{P}_t  \approx \mathcal{N}(\hat{\theta}_2,....,\hat{\theta}_K; \hat{\Sigma})
\end{split}
\end{equation}
We can efficiently solve the maximization problem in \eqref{maximizer} using any smooth convex optimization algorithm (for example Newton's method). Here, the Hessian is readily available given the form of the likelihood, and so approximate sampling of the posterior is computationally efficient.  

Now we can proceed to formulate our algorithms. Given the posterior sampling scheme described above, there are two popular ways of running Thompson Sampling. The canonical way \citep{thompson1933likelihood, chapelle2011empirical} samples parameters from the current posterior distribution and takes the action that maximizes the expected reward, which aims at achieving low regret \citep{pmlr-v23-agrawal12, Agrawal2017, Kaufmann2012ThompsonSA, Russo2016AnIA}. In our setting, we sample $\theta_2,..,\theta_K$ using \eqref{maximizer} and choose the vaccine group with the maximum $\theta$. Algorithm \ref{algo:PLTS} gives the details.

In our experiments, we also consider a PLTS-based adaptation of the
top-two Thompson Sampling algorithm of \citet{russo2020simple}.
This adaptation, where the second best action is selected in any given round with fixed probability $\beta$, targets the best arm identification problem. Algorithm \ref{algo:TTPLTS} details the algorithm. The algorithm takes in a parameter $\beta$ which indicates the probability of sampling from the second best arm. We fix $\beta = 0.5$, which \citet{russo2020simple} suggests as a safe default choice. 

\setlength{\textfloatsep}{12pt}
\begin{algorithm}[htb]
 \floatname{algorithm}{Algorithm}
    \caption{\label{algo:PLTS} 
             Partial Likelihood Thompson Sampling}
    \begin{algorithmic}
        \For{$t = 0,...,T-1$}
             \If{no infection has happened to any arm}
                  \State Assign new participants uniformly randomly.
             \Else 
                  \State Update posterior using (\ref{maximizer}).
                  \For{each $m_t$ newly arrived participant}
                  \State Sample $(\theta_2,...,\theta_K)$ and set $\theta_1 = 0$. 
                  \State Assign participant to group $\arg\max_{i} \theta_i$. 
                  \EndFor
             \EndIf
        \EndFor
    \end{algorithmic}
% \vskip -0.1in
\end{algorithm}

\setlength{\textfloatsep}{12pt}
\begin{algorithm}[htb]
\floatname{algorithm}{Algorithm}
    \caption{\label{algo:TTPLTS} 
             Top Two Partial Likelihood Thompson Sampling ($\beta$)}
    \begin{algorithmic}
    \For{$t = 0,...,T-1$}
        \If{no infection has happened to any arm}
            \State Assign new participants uniformly randomly.
        \Else 
            \State Update posterior using (\ref{maximizer}).
            \For{each $m_t$ newly arrived participant}
                  \State Sample $(\theta_2,...,\theta_K)$ and set $\theta_1 = 0$. 
                  \State $I \leftarrow \arg\max_{i} \theta_i$.
                  \State Sample $B \sim \text{Bernoulli}(\beta)$.
                  \If{$B = 1$}
                       \State Assign participant to group $I$.
                  \Else
                        \State \textbf{repeat}
              \State \,\,\, \,\,\, Sample $(\theta_2,...,\theta_K)$ and set $\theta_1 = 0$. 
              \State \,\,\, \,\,\, $J \leftarrow \arg\max_{i} \theta_i$.
              \State \textbf{until} $J \ne I$  
              \State Assign participant to group $J$.
                  \EndIf
            \EndFor
        \EndIf
    \EndFor
 \end{algorithmic}
 % \vskip -0.1in
\end{algorithm}

\section{Evaluation using COVID-19 Infections Data} 
\label{sec: experiments}

We conduct semi-synthetic experiments in this section which is motivated by the real COVID-19 vaccines and case counts. Specifically, we model the baseline hazard rate using real world COVID-19 infections data and choose the efficiency parameters according to the efficacy of some approved vaccines. We evaluate our algorithm in both tasks, getting as few infections as possible and correctly identifying the best vaccine. 

We simulate our experiment using Protocol \ref{protocol:trial_discrete} and fix our length of experiment to be $T = 200$. For simplicity we let the number of new participants to be constant at each time step, i.e. $m_t$ is a constant. Denote the total number of volunteers by $M$, we let $m_t = \frac{M}{T}$.

\subsection{Data}

\begin{figure}[t]
    \centering
    \includegraphics[width=0.7\columnwidth]{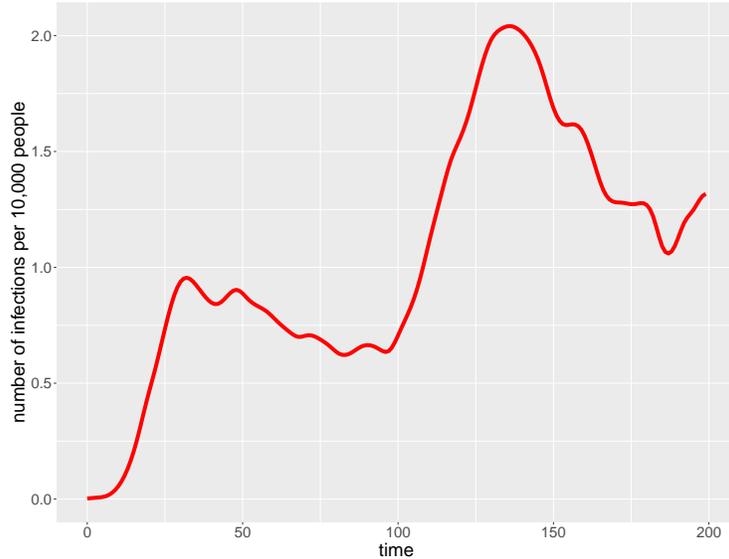}
    \caption{Baseline hazard rate as a function of time using moving average daily infections data provided by the CDC.}
    \label{fig:baseline_hazard}
\end{figure}

To model the baseline (or placebo) hazard rate, we use the data of 7-day moving average infections in US provided by the CDC data tracker \citep{CDC}. We pick the period of 200 days starting from March 9th, 2020. To get $h_1(t)$ we divide the daily infection numbers by the US population. The resulting baseline hazard rate is shown in Figure \ref{fig:baseline_hazard}. We clearly see two distinct waves of infections that occurred during this 200-day period.
 
Our next task is to set the efficiency parameters $\theta_k$ corresponding to the non-placebo
study arms. To do so, we use point estimates from a number of randomized controlled trials run early
in the COVID-19 pandemic. Specifically:
\begin{itemize}
\item Based on AstraZeneca Vaccine trials with a 70\%
reported efficacy \citep{VOYSEY2021881}, we set $\theta_2 = 1.2$.
\item Based on SinoPharm Vaccine trials with a 78\%
reported efficacy \citep{sinopharm}, we set $\theta_3 = 1.5$.
\item Based on Novavax Vaccine trials with an 89\%
reported efficacy \citep{Novavax}, we set $\theta_4 = 2.2$.
\item Based on Sputnik Vaccine trials with a 91\%
reported efficacy \citep{LOGUNOV2021671}, we set $\theta_5 = 2.4$.
\item Based on Pfizer and Moderna Vaccine trials with roughly 95\%
reported efficacies \citep{doi:10.1056/NEJMoa2034577,doi:10.1056/NEJMoa2035389}, we set $\theta_6 = 3.0$.
\end{itemize}
The motivation for using these numbers is that we hope they capture realistic
effect sizes one might see in a multi-arm vaccine trial, and not necessarily that
they exactly match real-world efficiencies of the above vaccines established after
pooling data from multiple trials.

\subsection{Evaluation Metrics}

We evaluate the performance of each experimental design using the following metrics;
throughout, we use the fact that the 6-th arm is best to condense notation.
\begin{itemize}
\item In-sample regret (ISR): Defined as $\frac{1}{T}\sum_{t=1}^T \theta_6-\theta_{I_t}$ where $I_t$ is the action chosen at round $t$.
\item Best arm identification probability (BIP), i.e., the fraction of times that the best arm (here, the 6-th)
has the lowest estimated infection hazard. Specifically, let $A_i$ be the estimated best arm for replication $i$, the best arm identification probability is defined as $\sum_{i=1}^{B} \mathbf{1}\{A_i = 6\}/B$.
\item Expected policy regret (EPR), as defined in \citet{exploration_sampling}: Let $a$ be the estimated best action, let $\Delta_a = \theta_6-\theta_a$. This is defined as 
\begin{equation}
\label{eq:policy_regret}
     \sum_{i=1}^{5} \frac{\sum_{j=1}^{B} \mathbf{1}\{A_j = i\}}{B}\Delta_a
\end{equation}
\end{itemize}
Of these metrics, the first measures the ``cost'' of running the experiment (i.e., how many study
participants were assigned to suboptimal arms during the trial), while the latter two measure the
quality of the findings from the study.

\subsection{Methods Under Comparison}

Our goal is to evaluate our proposed method, PLTS,
as well as the top-two Thompson sampling based variant designed for best-arm
identification (TTPLTS). We compare these methods to two baselines:
A standard, uniformly randomized controlled trial (RCT), and the
delayed exponential weighting (DEW) algorithm of \citet{Thune2019NonstochasticMB}.

\setlength{\textfloatsep}{12pt}
\begin{algorithm}[t]
 \floatname{algorithm}{Algorithm}
    \caption{\label{algo:dew} 
             Delayed Exponential Weights (DEW)}
    \begin{algorithmic}
        \State \textbf{Input:} Learning rate $\eta$, number of arms $K$
        \State Initialize weights $w_0^a = 1, \,\, \forall a = 1,...,K$
        \For{$t=1,..,T$}
             \State Let $p_t^a = \frac{w_t^a}{\sum_b w_t^b}$ for $a = 1,...,K$
             \State Place newly arrived volunteers according to distribution $\mathbf{p_t}$
             \State Observe set of infections $(s,a)$ where $s$ is the time of enrollment and $a$ is the vaccine group
             \State For each infection $(s,a)$, let $w_t^a = w_{t-1}^a\exp(-\eta\cdot\frac{1}{p_s^a})$
        \EndFor
    \end{algorithmic}
% \vskip -0.1in
\end{algorithm}

As discussed in the related work section, we are not aware of any existing
methods for sequential experimentation that were designed for our setting, i.e.,
with only negative feedback, unbounded delays in receiving feedback, and time-varying
delays. However, the DEW approach, although introduced and studied in an adversarial
setting with bounded delays, is simple and flexible enough that---at least
algorithmically---it can be used in our setting, which is why we also explore using it as a baseline.

The DEW algorithm is a form of exponential weighting where weights are updated
whenever negative rewards are observed; see Algorithm \ref{algo:dew} for details.
The one major challenge in using this algorithm is in choosing the learning rate $\eta$.
\citet{Thune2019NonstochasticMB} offer guidance based on bounds on the delay distribution,
but here of course we have no such bounds (and our setting does not fall under the purview of
their theory), so it wasn't clear to us how to choose $\eta$. Thus, we simply consider
3 baselines, DEW with $\eta = 0.01, 0.1, 0.4$, that span the range of behaviors one can
get from the method. (For Thompson sampling, we use an uninformative prior and so there
is no analogous tuning parameter for the learning rate that needs to be specified.)

Finally, in order to evaluate best arm identification probabilities, we need each method to output a recommended best arm at the end of the experiments. PLTS and TTPLTS output the vaccine with the largest posterior mode and DEW outputs the vaccine with the largest weight. RCT picks the vaccine with lowest infection rate at the end of the trial.

\subsection{Results}

For each method we consider, we use a sample size of $M = 60000$ study participants 
and replicate all simulations 1000 times. Table \ref{tab:simulation_volunteer} summarizes
the results across all methods and performance metrics.

\begin{table}[t]
\centering
\begin{tabular}{|c|ccc|}
 \hline
   & \multicolumn{3}{c|}{Metric} \\
   Method & ISR & BIP (\%) & EPR \\ 
    \hline
   RCT & 385.08 (0.04) & 86.0 (1.1) & 0.090 (0.007) \\ 
   $\eta=0.01$ & 297.14 (7.71)& 86.6 (1.1) & 0.086 (0.007) \\ 
   $\eta=0.1$ & 186.03 (24.45)& 89.7 (1.0) & 0.067 (0.006) \\
   $\eta=0.4$ & 158.23 (39.80)& 79.3 (1.3) & 0.149 (0.010) \\
   PLTS & 160.25 (0.96)& 91.8 (0.9) & 0.052 (0.006) \\
   TTPLTS & 183.76 (0.71)& 93.5 (0.8) & 0.041 (0.005) \\ 
   \hline
\end{tabular}
\caption{Results comparing Partial Likelihood Thompson sampling (PLTS), top-two PLTS (TTPLTS), DEW with varying learning rate ($\eta=0.4, 0.1, 0.01$) and the randomized controlled trial (RCT). We display three metrics defined previously and fix $M = 60000$. Standard errors are given in parentheses; each configuration is replicated 1000 times.}
\label{tab:simulation_volunteer}
\end{table}

\begin{figure}[t]
    \centerline{\includegraphics[width=0.7\columnwidth]{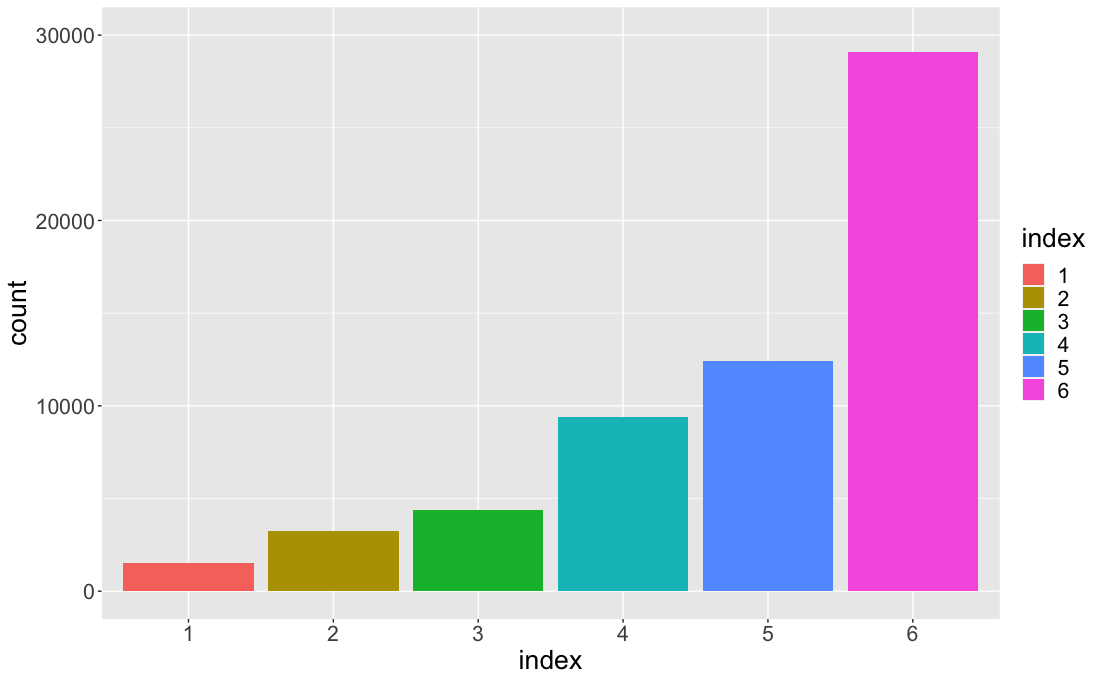}}
    \caption{ Total number of participants assigned to each vaccine group for the method PLTS with $M=60000$ total participants, averaged over 1000 replications.}
    \label{fig:total_share}
\end{figure}

Our first comparison is between the simplest variant of our methods, PLTS, and the
RCT baseline. We here see that PLTS outperforms the RCT along all metrics: It both
achieves smaller in-sample regret and has more power to identify the best arm. The reason
it can do so is that it quickly shifts sampling towards the most promising vaccines;
see Figure \ref{fig:total_share}.
This is clearly desirable from a regret minimization point of view, but here it is
also desirable from a power point of view since it concentrates sampling on the most
difficult questions, i.e., distinguishing the best arms from each other.

\begin{figure}[htb]
    \centerline{\includegraphics[width=0.9\columnwidth]{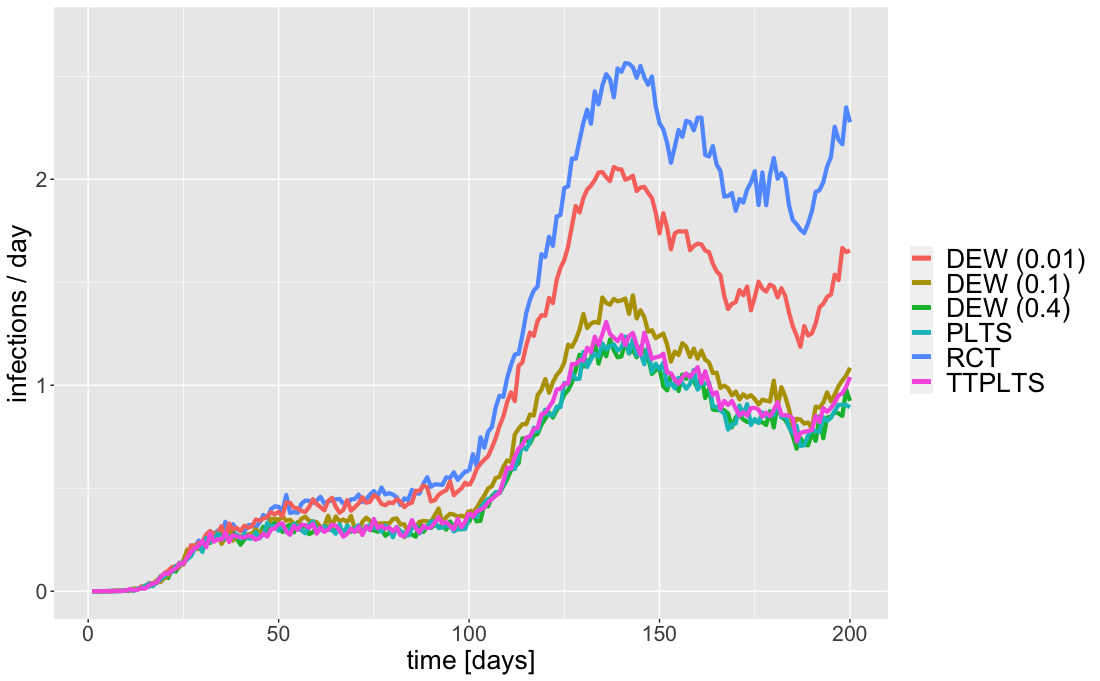}}
    \caption{Total infections across all vaccine groups for 6 methods we consider: DEW with varying learning rate ($\eta=0.4, 0.1, 0.01$), RCT, PLTS and TTPLTS, averaged over 1000 replications with total number of participants $M=60000$.}
    \label{fig:infection}
\end{figure}

\begin{figure*}[htb]
        \centering
        \begin{subfigure}[b]{0.475\textwidth}
            \centering
            \includegraphics[width=\textwidth, ]{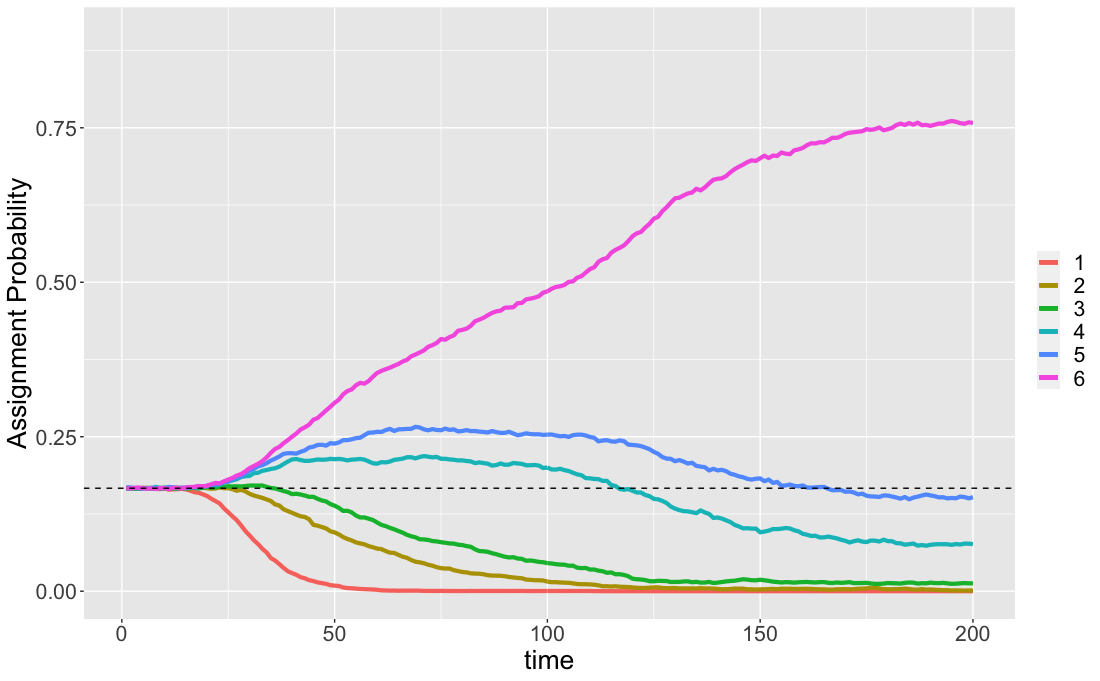}
            \caption[]%
            {{\small DEW with $\eta = 0.4$}}    

        \end{subfigure}
        \hfill
        \begin{subfigure}[b]{0.475\textwidth}  
            \centering 
            \includegraphics[width=\textwidth]{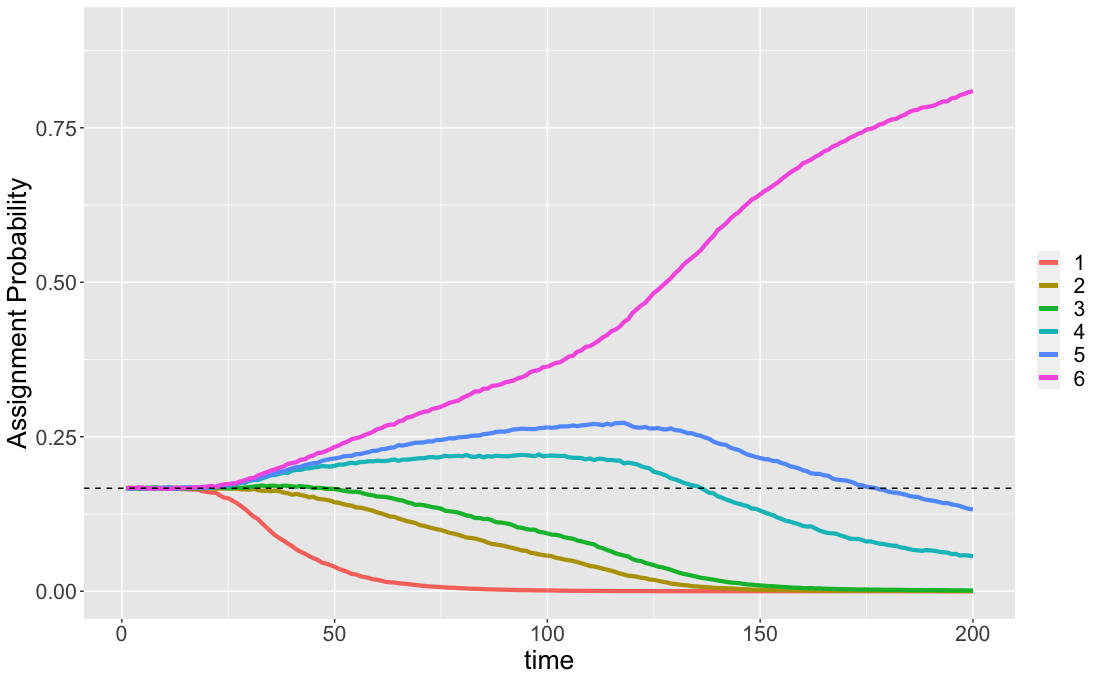}
            \caption[]%
            {{\small DEW with $\eta = 0.1$}}    

        \end{subfigure}
        \vskip\baselineskip
        \begin{subfigure}[b]{0.475\textwidth}   
            \centering 
            \includegraphics[width=\textwidth]{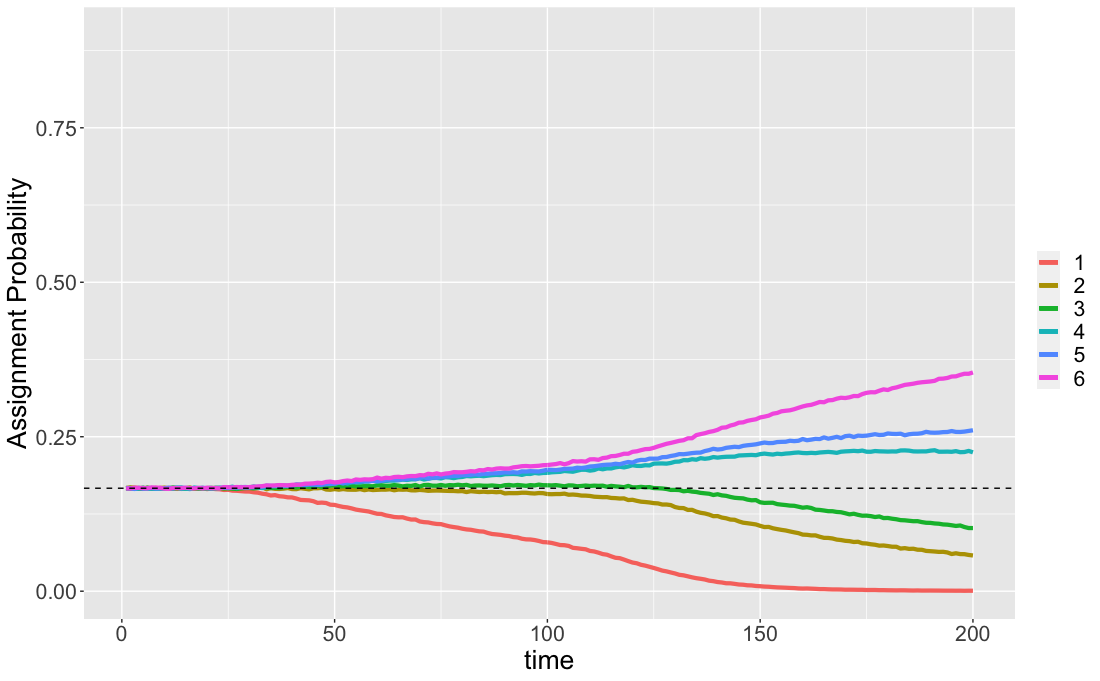}
            \caption[]%
            {{\small DEW with $\eta = 0.01$}}    
        \end{subfigure}
        \hfill
        \begin{subfigure}[b]{0.475\textwidth}   
            \centering 
            \includegraphics[width=\textwidth]{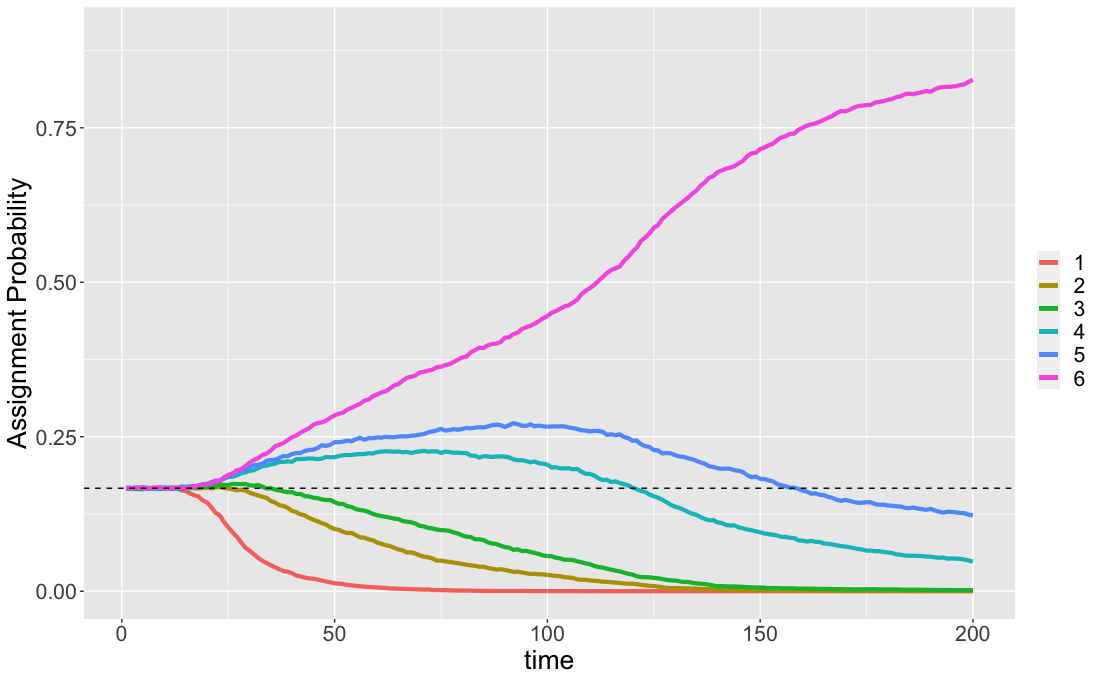}
            \caption[]%
            {{\small PLTS}}    
        \end{subfigure}
        \caption[ ]
        {\small Plot of assignmnet probability of each vaccine group as a function of time for DEW with varying learning rate $\eta$ and PLTS when $M = 60000$, averaged over 1000 replicates. } 
        \label{fig:prob}
\end{figure*}

Next, we compare PLTS to the DEW baselines in terms of in-sample regret. Here, the
picture is nuanced. When well tuned, DEW can slightly outperform PLTS; however, it
is not clear whether an adaptive tuning parameter choice could mirror this result.
The rate at which all methods incur infections during the study is shown in Figure
\ref{fig:infection}. We see that all the methods incur similar numbers of infections
in the first wave, but the well-performing methods are able to focus on the better
arms and considerably cut down on infections by the time we get to the second (larger)
wave.\footnote{One reason all study designs still have a fairly large number of infections in
the second wave is that all designs assigned a non-trivial number of subjects to the
less effective arms, including the placebo, at the beginning of the study, and that
many of these participants were then vulnerable to infection in the second wave.}

The comparison looks different, however, once we look at metrics that consider the quality
of the selected arm, i.e., best arm identification probability and policy regret. Here,
PLTS still does well, but variants of DEW that achieved small in-sample regret do very poorly.
It appears that, in order to achieve good in-sample regret, DEW needs to make unstable
or greedy choices that hurt the quality of the selected arm. In contrast, PLTS is able to
focus on the best arms without suffering from this phenomenon.

Relative to PLTS, the top-two variant TTPLTS achieves better post-trial metrics but
worse in-sample regret. This is to be expected, since TTPLTS invests more in sampling
the second-best arm in order to improve power for best arm identification. Whether
a practitioner prefers the behavior of PLTS or TTPLTS will depend on the relative
importance they give to in-sample versus post-trial performance metrics.

Finally, we investigate how arm-assignment probabilities of different methods evolve over time:
Figure \ref{fig:prob} shows the assignment probabilities averaged over 1000 replicates for each vaccine candidate as a function of time for both DEW and PLTS. The dashed horizontal line shows the uniform probability RCT uses. We see that in both cases the more promising candidates get larger shares as time goes on. However, we do see that when DEW uses a large learning rate (corresponding to the cases with
good in-sample regret), the assignment probabilities almost flatten out as we approach the end of the
trial, suggesting that by this point the learning rate has become too fast to enable reliable convergence to the best arm.

\section{Discussion}

Sequential experiments have considerable potential to address challenges associated by
new disease variants that emerge during a pandemic \citep{castillo2021market}. However,
the vaccine trial setting comes with a number of statistical challenges---including unbounded and time-varying delay
distributions and partial feedback---that have not been considered in the context of
existing bandit algorithms.
We introduced partial likelihood Thompson sampling, which
adapts Thompson sampling to the setting of vaccine trials using fundamental modeling
techniques that have been prevalent for decades in the survival analysis literature \citep{cox1984analysis}.
We find our method to be a robust and performant option for sequential experimentation
in an experiment built around data from the COVID-19 pandemic, thus highlighting its
promise as a tool for quickly targeting the use of existing vaccines against a new disease variant. Additionally, our method can also be applied in other settings where the proportional hazards model is relevant; for example, in email marketing or online advertising. Our method deals with the complicated delay structure arising from these applications from a modeling perspective, thus opening the door for more efficient learning. 

\bibliographystyle{plainnat}
\bibliography{references}
\end{document}